# The Coupling of a Wigner Polaronic Charge Density Wave with a Fermi Liquid arising from the Instability of a Wigner Polaron Crystal: A possible Pairing Mechanism in High $T_c$ Superconductors.

A. Bianconi, M. Missori

*University of Rome "La Sapienza", Department of Physics, P. A. Moro 2, 00185 Roma, Italy*

**Summary.** We report evidence that the metallic phase in high $T_c$ cuprate superconductors results from an instability of the quasi 2D-electron gas at densities lower or larger than the critical density $\delta_c$ for the formation of a Wigner polaronic crystal. At $\delta_{tot} \neq \delta_c$ the electron gas undergoes a microscopic phase separation between 1) a polaron gas with density $\delta_l$ at the *Wigner localization limit* and 2) a *Fermi liquid confined in a superlattice of quantum stripes* with density $\delta_i$. The 2D-polaron gas condenses into a paired unidimensional Wigner charge density wave (CDW) at $T<T^* \sim 1.4\ T_c$. Below the superconducting critical temperature the two types of carriers are coupled, and the pairing mechanism involves virtual polaron pairs. This mechanism drives the condensate to the maximum critical temperature for a Fermi liquid. In addition the pairing model proposed here correctly predicts the dependence of $T_c$ on the condensate density $n_s/m^*$ (Uemura plot), and the variation of the isotope effect from the weak coupling limit a~0.5 where $\delta_i < \delta_l$ to $\alpha \sim 0.1$ where $\delta_i > \delta_l$.

1. INTRODUCTION

Starting from the recent experimental determination [1] of the polaron area $S_p$ we present a new model a) for the quantum state of the electron gas in the cuprate superconductors and b) for the microscopic pairing mechanism. The quantum state of the electron gas in the metallic phase is described by a microscopic phase separation between a Fermi liquid and a polaron gas close to the Wigner localization limit. This phase separation results from the instability of the Wigner polaron gas for densities close but different from the critical density for a generalized Wigner polaronic crystal. The pairing mechanism results from the coupling between the two types of charge carriers.

We want to show here that the present model allows us to understand some key phenomenological points of the physics of cuprate superconductors that up to now remained obscure:

1) the insulator to metal transition at finite doping [2];

2) the electron lattice instability at the critical doping 1/8 [3];

3) the anomalous normal metallic phase;

4) the opening of the spin gap at the temperature $T^* > T_c$ [4];

5) the variation of the exponent $\alpha$ of the isotope effect from the weak coupling value $\alpha \sim 0.5$ at low doping and $\alpha < 0.1$ at optimum doping [5];

6) the relation of $T_c$ versus $n_s/m^*$, known as Uemura plot, where $n_s$ is the condensate density and $m^*$ is the effective mass [6].

As shown in some recent works [1,7-9] the metallic phase of the electron gas in the cuprate superconductors results from a microscopic phase separation between a polaron gas and a Fermi liquid. By doping the antiferromagnetic insulator with a total number of holes per Cu site $\delta_{tot}$ two types of holes are formed: a polaron gas with average hole doping $<\delta_l>$ and a doped spin fluid forming a Fermi liquid with an average hole doping $<\delta_i>$.

The phases of the quantum state of the electron gas in cuprate high $T_c$ superconductors as function of doping are described by the polaron pair filling factor $\nu = 2\rho S_p$, where $\rho$ is the total surface hole density in the $CuO_2$ plane. The homogeneous phases of the polaron gas are formed at integer values of the polaron pair filling factor:

1) (a) $\nu=1$ indicates the formation of a generalized Wigner [10] charge density wave (CDW) where the polarons cover half of the surface of the $CuO_2$ plane. (b) the 1/8 instability [3] is shown to be due to the formation of a generalized Wigner [10] polaronic crystal, where the polarons cover the full surface of the $CuO_2$ plane, i. e. $\nu=2$.

2) In the insulator phase at very low doping, defined by the polaron filling factor in the range $\nu < 1$, the polaron gas is in the Wigner low density insulating phase in fact we demonstrate that its density parameter $r'_s > 37$ [10] and the insulator to metal transition takes place at $\nu=1$.

3) The metallic phase is assigned to the instability of the electron gas at non integer values of $\nu$. The anomalous normal metallic state is due to the coexistence of a polaron gas condensed into a polaronic generalized Wigner charge density wave, where the local doping is $\delta_l=1/8$ and a Fermi liquid with local doping $\delta_i <1/8$ or $\delta_i >1/8$ at the metallic low or optimum doping regime respectively.

*The nature of the microscopic pairing mechanism at $T_c$ is revealed by the following experimental results:*

4) The opening of the spin gap [5] at the temperature $T^* > T_c$ corresponds to the critical temperature $T^*$ for the formation of the self trapped polarons [1] and therefore it is assigned to the spin pairing of polarons.

5) The superconductivity shows two doping regimes defined by the relation $\delta_i < \delta_l$ and $\delta_i > \delta_l$ where the exponent of the isotope effect is $\alpha \sim 0.5$ and $\alpha \sim 0.1$ respectively.

6) The condensate with density $n_s$ is formed by *a portion of the Fermi liquid coupled with the polaron pairs*, where the polaron pairs play the role of virtual excitations. On the other hand the condensate can be also described as *polaron pairs which decay into superconducting electron pairs of the Fermi liquid*. Therefore the portion of the electron gas which condenses will satisfy the condition that the coherence length is the same as the interatomic distance $\xi_0 \sim 2r_s a_B$ that is valid for an electron gas close to the Wigner crystallization [11, 12], where $a_B$ is the Bohr radius and $r_s$ the electron density parameter. This condition is close to the condition giving the maximum critical temperature for a Fermi liquid $k_F \xi_0 \sim 2\pi$ [13,14]. We show that this pairing mechanism correctly predicts the experimental dependence of $T_c$ versus $n_s/m^*$.

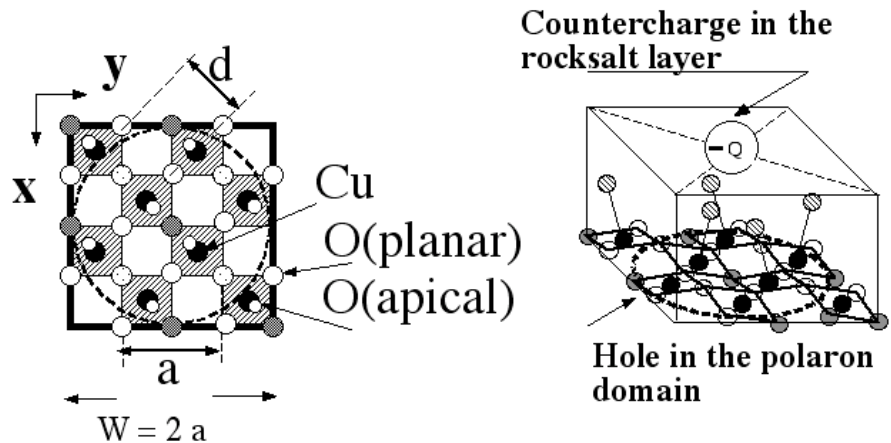

**Fig. 1.** Pictorial view of a single polaronic state where the hole L(a) in the $CuO_2$ plane with mixed $b_1$, $a_1$ and e symmetry trapped in a polaronic domain of width W=2a including 8 Cu sites. In the left picture the hole is trapped in a polaronic impurity state by the negative counterions on the BiO layer.

## 2. THE POLARON SIZE AND ITS TEMPERATURE DEPENDENCE

There are many experimental data on cuprate superconductors showing that the holes (or electrons) in the $CuO_2$ plane induced by doping can form both 1) a polaronic charge [15-19] and

2) itinerant charge carriers forming a Fermi liquid [20-23].

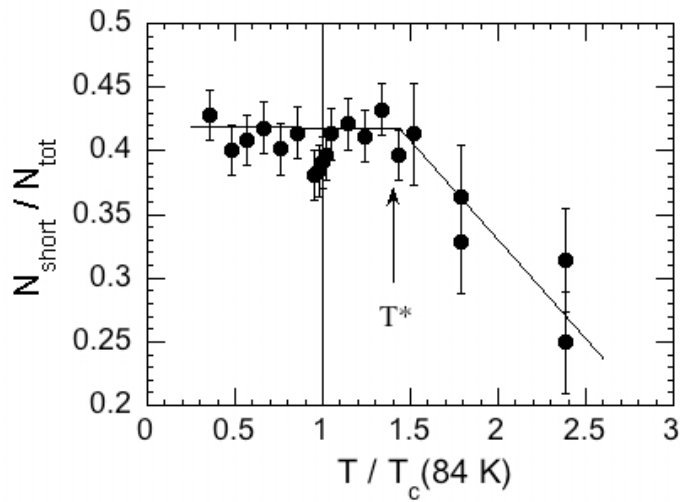

**Fig. 2.** The measured values of the effective number of short Cu-O(apical) bonds $N_{short}/N_{tot}$ as function of temperature in a monodomain crystal Bi2212 and the values of the width of the polaron stripe $W/a = (W/\lambda_p) \lambda_p/a = (N_{short}/N_{tot}) \lambda_p/a$. Where $\lambda_p/a$ is 4.65 as measured by electron diffraction. The width W at $T < T^* = 120K \sim 1.4 T_c$ is $W = 10.54 \pm 0.6$ Å.

The presence of polaronic charges is well established in the insulating phase at very low doping regime, on the contrary the presence of itinerant charges forming a correlated Fermi liquid with a large Fermi surface is well established in the optimum doping regime.

To solve the puzzle of the quantum state of high $T_c$ superconductivity we have performed two key experiments to determine 1) the size of the polaron and 2) the ordering of the polaron gas in the metallic phase. In these experiments we have measured 1) the polaron size in a $Bi_2Sr_2CaCuO_{8+y-}$ (Bi 2212) crystal at optimum doping (The total hole doping was about 0.18 holes per Cu site in our crystal) by EXAFS and 2) the ordering of the polaronic impurity states by diffraction.

We have focused our interest in $Bi_2Sr_2CaCuO_{8+y}$ because is the ideal case to study the non homogeneous $CuO_2$ plane. In fact in this system the acceptors out of plane (additional oxygen in the BiO plane) are ordered along lines in the a-axis direction in the metallic phase. On the contrary the ordering of the acceptors takes place in small domains in most of the families of cuprate superconductors and it cannot be observed by most of the available experimental methods.

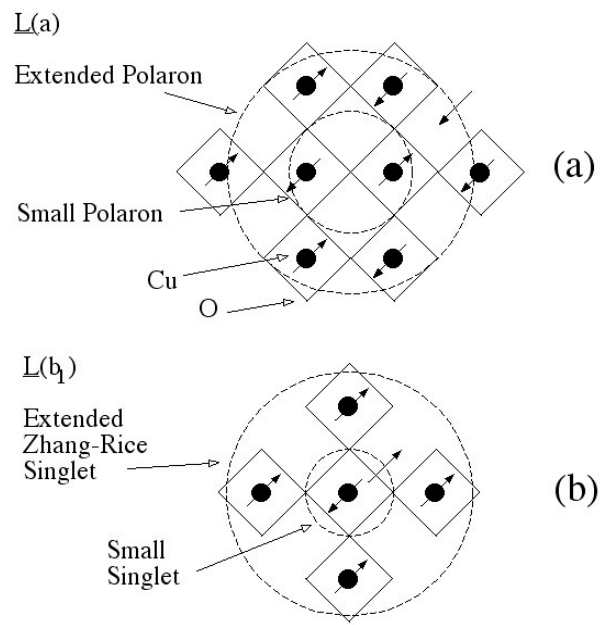

**Fig. 3.** Pictorial view of the polaron found in this work, panel (a), extended over an even number of Cu sites associated with L(a) holes in orbitals with mixed symmetry, and of the extended and small Zhang Rice singlets $3d^9\underline{L}(b_1)$ on a odd number of Cu sites, panel (b). The small and extended hole states $\underline{L}(a)$ or $\underline{L}(b_1)$ added by doping move in the antiferromagnetic lattice of the parent compound.

The polaronic domain that we have found is shown in Fig. 1. It can be described by a conformational coordinate the Cu-O(apical) distance which becomes shorter, mainly due to the displacement of the Cu ion out of plane, and the tilting of the $CuO_4$ square planes. The $CuO_2$ plane has been shown to be decorated by stripes of polarons associated to holes trapped by the acceptors in linear arrays [7-9] at optimum doping. Two Cu-O(apical) distances 2.40±0.02 Å and 2.57±0.02 Å below 120K have been found by polarized EXAFS, in agreement with diffraction works [24] confirming the presence of domains with different Cu site configurations in the $CuO_2$ plane. We have measured the width W of the polaronic stripes by measuring the ratio between the number $N_{short}$ of short Cu-O(apical) bonds and the total number $N_{tot}$ of the Cu-O(apical) bonds in a monodomain oriented crystal by polarized EXAFS.

In Fig. 2 we have obtained W by the measured values of $N_{short}/N_{tot}$ and by measuring the superstructure period $\lambda_p$ by electron diffraction

$$W/a = (W/\lambda_p) \lambda_p/a = (N_{short}/N_{tot}) \lambda_p/a$$

The width W is about a=5.4 Å at room temperature in agreement with neutron diffraction data [24] and it becomes larger at T< $T^*$=120K~1.4$T_c$ where it is W=10.54 ± 0.6 Å. Therefore W~2a =10.8 Å, which is the side of the square domain shown in Fig. 1, and we assign $T^*$ to the temperature for the polaron self trapping in fact W ~ 2a is constant below $T^*$ except for the anomaly around $T_c$.

## 3. THE POLARON PAIR FILLING FACTOR AND THE INSULATOR TO METAL TRANSITION

The structure of the polaronic domain shown in Fig. 1 is characterized by $CuO_4$ square planes with a LTT-type distortion i.e. similar to the low temperature tetragonal phase (LTT) observed in $La_{1.875}Ba_{0.125}CuO_4$ [3] and a displacement of the Cu ion out of plane, giving a short Cu-O(apical) bond. These two set of distortions of the $CuO_5$ pyramid introduce a mixing of the molecular orbitals of $b_1$ symmetry ($3d_{x^2-y^2}$ O(planar) $2p_{x,y}$), with orbitals of $a_1$ {Cu $3d_{3z^2-r^2}$ O(planar) $2p_{x,y}$, O(apical)$2p_z$} and orbitals with e {Cu $3d_{xz}$ $3d_{yz}$ O(planar)$2p_z$} symmetry forming new molecular orbitals with larger components of the charge density out of plane giving the polaronic hole that we denote as $\underline{L}(a)$. For these reasons this polaron was called a $3d_{3z^2+r^2}$ polaron [7]. The polaron domain, shown in Fig. 1 is extended over an even number of Cu sites, which confirms that it is associated with a hole of symmetry different from the holes of $b_1$ symmetry forming the antiferromagnetic lattice in the parent compound (see Fig. 3a). A polaron associated with a ligand hole $\underline{L}(b_1)$ of same symmetry as the holes of $b_1$ symmetry forming the antiferromagnetic lattice in the parent compound, i. e. a Zhang Rice singlet is expected to be centered on a Cu ions and extended over an odd number of Cu sites (see Fig. 3b). The polaron effective mass $m_p = m/m_e$, where $m_e$ is the electron mass, can be estimated from the polaron radius to be about 16.

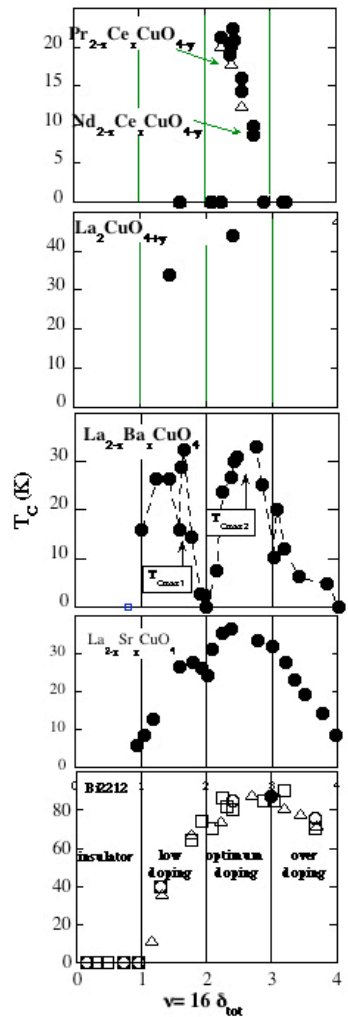

**Fig. 4.** $T_c$ versus the polaron pair filling factor for Bi2Sr2Ca1-xYxCu2O8+y system as measured by Quitmann et al. (triangles), Groen et al. (squares) and by Pettiti et al. (circles). $T_c$ versus $\nu$ for hole doped LaSrCuO, LaBaCuO, La2CuO4+y and electron doped systems.

The polaron area is found to be $S_p=4a^2$ (where a=5.4 Å). Usually the doping in the cuprates is measured as the number of holes per Cu site $\delta_{tot}$. The surface density is given by $\rho=\delta_{tot}/d^2$ where d is the Cu-Cu distance, about 3.8 Å, therefore $\rho=2\delta/a^2$ where $a= d \sqrt{2}$. Having determined the polaron area we can now measure the hole density by the polaron pair filling factor which is given by

$$\nu = 2\, \rho.\, Sp = 16\, \delta_{tot}.$$

We have reported in Fig. 4 the critical temperature of the $Bi_2Sr_2Ca_{1-x}Y_xCu_2O_{8+y}$ system as function of the polaron pair filling factor [25-27], and it is compared with the similar curves for other superconducting families. We clearly show that 1) the system is insulator for $\nu<1$, the metal insulator transition takes place at the critical integer number $\nu=1$ which corresponds to the critical density for percolation in two dimensions, suppressing the long range antiferromagnetic order in the undistorted lattice outside the polaron domain walls, and forming a spin liquid. At the critical value of $\nu=1$, where the polaron gas covers half of the plane surface the ordering of the polarons driven by the short range attraction and long range Coulomb repulsion is expected. If the Coulomb repulsion overcomes the kinetic energy at this critical density a generalized Wigner charge density wave could appear as will be show below. The plots of $T_c$ versus the filling factor for all superconducting families shown that the insulator to metal transition occur always at $\nu=1$. This result clearly supports our hypothesis that the size of the polaron is a characteristic property of the $CuO_2$ plane and therefore it is the same in all families of cuprate superconductors.

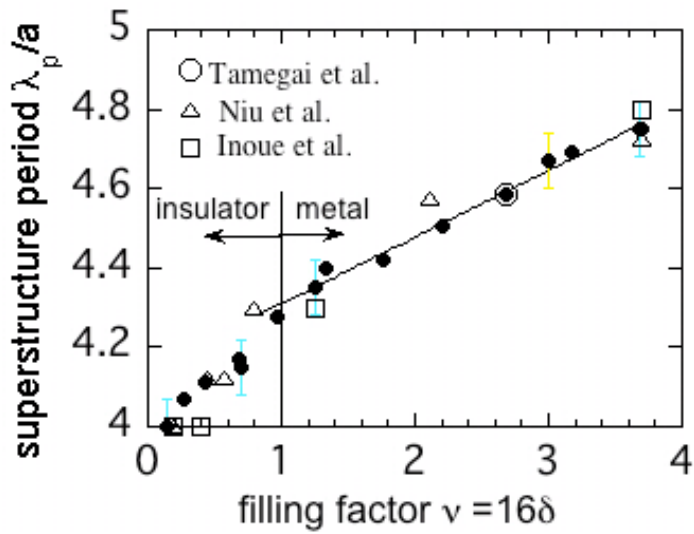

**Fig. 5**. The superstructure period $\lambda_p/a$ of $Bi_2Sr_2Ca_{1-x}Y_xCu_2O_{8+y}$ as function of the doping measured by the polaron pair filling factor $\nu$. The solid line is the fitting in the range of the metallic phase for $\nu>1$.

## 4. ORDERING OF POLARONS AS FUNCTION OF DOPING

The ordering of the polarons along linear arrays in the metallic phase of Bi2212 has been already discussed in the first workshop on phase separation [9]. The polarons form parallel stripes resulting in a unidimensional charge density wave in the $CuO_2$ plane and its period is the same as the superstructure period of the full crystal. The incommensurate period $\lambda_p$ as function of doping has been measured by x-ray diffraction and electron diffraction in samples where the hole doping is controlled by Y doping $Bi_2Sr_2Ca_{1-x}Y_xCu_2O_{8+y}$. The superstructure period is 4a in the insulating case due to the misfit of the BiO and the $CuO_2$ planes but in the metallic phase it shows a linear behavior from $\lambda_p$=4.3a at the insulator to metal transition to about $\lambda_p$=4.8a in the over doped regime as shown in Fig. 5. The variation of the incommensurate superstructure period $\lambda_p$ as function of Y content reported in Fig. 5 in the *metallic phase* has been measured by several groups [28]. A correlation between $T_c$ and the superstructure period has been remarked by Pham et al. [29] in different heat treatments of the Bi2212 system. To obtain $\lambda_p$ as function of $\delta_{tot}$ shown in Fig. 5 we have fitted the data of the Y content x versus hole doping $\delta_{tot}$ reported by Groen et. al. [25], that is in good agreement with our experience on the characterization of this class of materials. Let us assume that the polaron size that we have measured is not doping dependent.

This can be justified in two ways. First, the polaron size is controlled by an intrinsic instability of the $CuO_2$ lattice for a structural transition from the LTO phase to the LTT phase where the Fermi energy is tuned at the van Howe singularity [30], which fixes the local density at 1/8 and therefore the polaron area. Second, in the metallic phase the polarons have condensed in a CDW therefore their local density is constant, $\delta_l$=1/8 as their distance W and their size, which is therefore constant in the metallic phase.

The experimental result that the ratio $N_{short}/N_{tot}$ is $\neq 0.5$ is a strong indication for the inharmonic electron lattice interaction giving an inharmonic CDW. The non-harmonic character can be associated with the electron lattice interaction at the critical local doping 1/8. In the presence of an inharmonic electron lattice interaction of pseudo Jahn Teller type [31] the inharmonic CDW drives the system toward a microscopic phase separation between polaronic holes L(a) condensed in linear arrays separated by a Fermi liquid formed by the holes L(b_1).

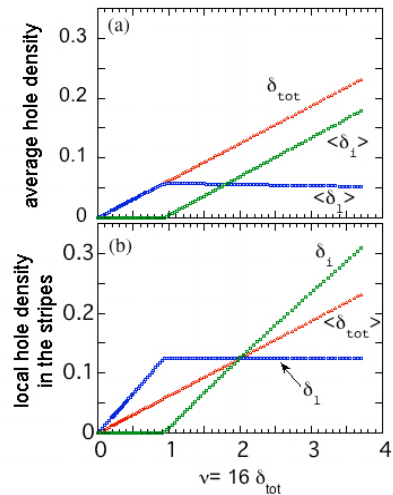

**Fig. 6.** Panel (a) The distribution of the total hole density $\delta_{tot}$ into the average hole density $\langle\delta_i\rangle$ of the itinerant electrons and the average hole density $\langle\delta_l\rangle$ of the polaronic charges. Panel (b) The local hole density $\delta_i$ of the itinerant electrons in the stripes of width L and the local hole density $\delta_l$ of polaronic charges.

## 5. PARTIAL CHARGE DENSITY FOR THE POLARON GAS AND THE FERMI LIQUID

Here we will show that starting from these two experimental data 1) Sp and 2) the relation between $\lambda_p$ and doping, we can solve the quantum state of the cuprate high $T_c$ superconductors.

In fact by starting from the experimental data of $\lambda_p(\nu)$ in Fig. 5 and by assuming that the polarons form stripes of width W=2a constant with doping, we can calculate the charge densities of the polaron gas and of the Fermi liquid as function of doping at T<T* shown in Fig. 6.

The average number of holes per Cu site $\langle\delta_l\rangle$ forming the polaron gas and the average number of holes per Cu site $\langle\delta_i\rangle$ forming the Fermi liquid are given by,

$$\langle\delta_l\rangle = a^2/(2\,W\,\lambda_p) = 1/(4\,\lambda_p/a) \text{ and } \langle\delta_i\rangle = \delta_{tot} - \langle\delta_l\rangle$$

and they are plotted in Fig. 6a. The local hole density $\delta_i$ in the stripes where the Fermi liquid is confined and the local hole density $\delta_l$ for the polaron gas are given by:

$$\delta_l = \langle\delta_l\rangle\,(\lambda_p/W) = 1/8 \text{ and } \delta_i = \langle\delta_i\rangle\,((\lambda_p/(\lambda_p-W)))$$

and they are plotted in Fig. 6b. In the metallic phase the polaron gas with average density $\delta_1$ is expected to give a small Fermi surface. The presence of a small Fermi surface of area $A=\pi k_F'^2 = 0.048\pm0.008$ Å$^2$, giving $k_F' \sim 0.123$ Å$^{-1}$ = 0.106 $(2\pi/a)$ has been measured by de Haas-van Alphen (dHvA) experiments [32] in YBa$_2$Cu$_3$O$_7$ (Y123) and Tl2201 systems. The incommensurate superstructure in Bi2212 with period $\lambda_p/a \sim 4.7$ and wave vector $q=2\pi/\lambda_p =0.21$ $(2\pi/a)$ at doping of about 18% is assigned here to a CDW driven by the nesting at the small Fermi surface for the average electronic structure in fact $q \sim 2 k_F'$.

A small Fermi surface pocket is expected to be at the S point but because the symmetry of the molecular orbital of these states, we think that this pocket actually appears in Bi2212. at the M $(\pi,0)$ point.

## 6. THE POLARON GAS CONDENSES IN A GENERALIZED WIGNER CDW AT ν=1 AND IN A GENERALIZED WIGNER CRYSTAL AT ν=2

The experimental research on the instabilities of the electron gas close to condense into a Wigner crystal [10] at a critical density of charge carriers $\rho < \rho_c$ has been at the basis of several unexpected phenomena in solid state physics in these last years. The formation of a Wigner crystal has been reported in the case of extreme low carrier concentration of the two dimensional (2D) electron gas on the surface of liquid helium [33]. The Wigner localization limit for a 2D-electron gas has been predicted at values of the electron density parameter $r_s>37$ where $\rho=1/\pi(r_s a_B)^2$, $a_B$ is the Bohr radius [10]. If an external field localizes the electrons, a 2D-electron gas can reach the Wigner localization limit at an electron density much higher than the expected $\rho_c$ for a free electron gas. In the quantum Hall effect [34, 35] the external perpendicular magnetic field confines the electrons to the magnetic length $l_c=(Ó/eB)^{1/2}$ i. e. within a surface $S=\pi Ó/eB$. The quantized anomalies in the Hall coefficient are observed at critical integer or rational numbers of filling factors $\nu = 2 S/(\pi (r_s a_B)^2) = 2 \rho S = \rho\, h/eB$.

Here we show that in the high $T_c$ cuprate superconductors the polaronic impurity state, trapping the holes within the polaron area $S_p$, acts as the external force to drive this 2D-polaron gas toward the Wigner localization limit. The polaron gas will reach the Wigner localization limit at the critical density $\rho_c = 1/S_p$. In these conditions the Coulomb repulsion between polarons overcomes the kinetic term and the condition $r_s' = r_s m_p/\varepsilon_{\tilde{Y}} > 37$ for Wigner localization of a 2D-electron gas is satisfied, where $r_s$ is the density parameter of the polaron gas of density $\rho_l$, and $m_p$ is the polaron effective mass. The phase diagram of the polaron gas will be described as in the quantum Hall effect by the polaron pair filling factor $\nu = 2\rho \cdot S_p$ which predicts the formation of homogeneous phases of the polaron gas at integer values of $\nu$: a Wigner charge density wave (CDW) at $\nu=1$, a the Wigner crystal at $\nu=2$, and a bipolaron crystal is expected at $\nu=4$, but it will be shown that it is forbidden by the Coulomb repulsion.

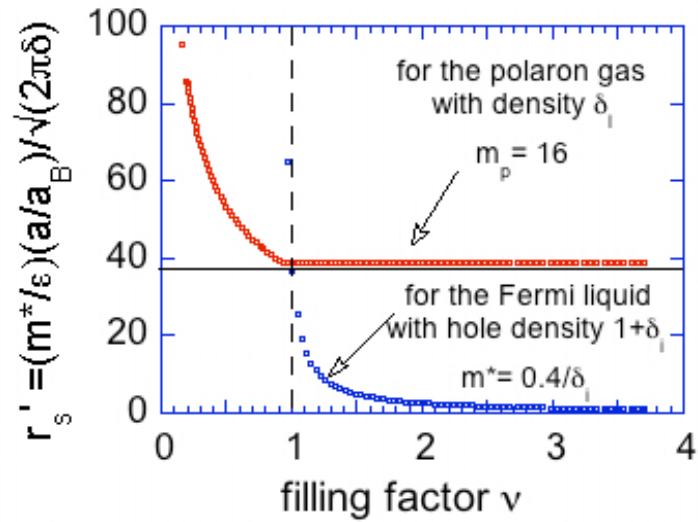

**Fig. 7.** The effective electron density parameter $r_s'$ for the polaron gas and for the Fermi liquid as function of doping.

In order to see if the polaron gas is in the Wigner localization limit we have calculated the effective polaron gas density parameter $r_s' = r_s m_p/\varepsilon_{\tilde{Y}}$ where $m_p$ is the polaron mass found to be here 16 and $1/\pi(r_s a_B)^2 = \rho_l = 2\delta_l/a^2$ is the polaron density. The dielectric constant $\varepsilon_{\tilde{Y}} = 4.75$ [36] has been used. The variation of $r_s'$, for the polaron gas shown in Fig. 7, clearly shows that it is always in the Wigner localization limit $r_s' > 37$ [10]. For the Fermi liquid we assume that the effective mass is $m^* = 0.4$ in absence of electron-electron correlations. Because the Fermi liquid results from doping a spin fluid with large electron-electron correlation giving an insulator at half filling we assume that the rinormalized effective mass is doping dependent following the standard expected law m* = 0.4 /$\delta_i$. The resulting effective density parameter for the Fermi liquid, shown in Fig. 7, shows that this liquid clearly is in the regime of an high density electron gas, except in a very narrow region close to the insulator-metal transition.

### 7. ON THE PAIRING MECHANISM

*We show here that the pairing mechanism arises from the coupling of a portion of the Fermi liquid with the Wigner polaronic CDW.* From Fig. 4 it is clear that the cuprates exhibit high $T_c$ superconductivity with a large Meissner fraction at two main critical hole (or electron) concentrations i. e. in the low doping region 1<ν<2 at ν ~ 1.8 where the maximum critical temperature is $T_{c1max}$ and in the optimum doping region 2<ν<3 at ν ~2.5-3 where the maximum critical temperature is $T_{c2max}$. These two phase can be described as resulting from the instability of the electron gas close to critical hole concentration for the polaron Wigner crystal at ν=2.

At $T_{c1max}$ the local density of the Fermi liquid $\delta_i$ in the Fermi liquid domain is smaller than the local density of the polaron gas $\delta_l$ in the polaronic domain, i. e. $\delta_l > \delta_i$, on the contrary at $T_{c2max}$ the opposite situation appears, where $\delta_l < \delta_i$.

The experimental evidence for the coupling between the Fermi liquid and the polaron gas giving superconducting pairs is shown by the fact that the optimum conditions for high $T_c$ superconductivity are realized at $T_{c1max}$ and $T_{c2max}$ where the ratio between the average density of two gas <$\delta_i$>/ <$\delta_l$> takes the integer values 1 and 2 respectively.

Let us consider the simplest case where <$\delta_i$> = <$\delta_l$>. This situation is realized for example in a single layer material ($La_2CuO_{4+y}$ with oxygen doping y of about 6% giving a superconducting material at $T_{c1max}$). For each additional interstitial oxygen in the rocksalt layer about two holes are injected in the $CuO_2$ plane, one hole is trapped in the polaronic impurity state and the second hole contributes to the Fermi liquid of free cariers.

It is interesting to remark that the charge neutrality of the metallic phase for the two different charge carriers can be imagined to be provided by the same oxygen atom out of the plane.

284

The pairing mechanism can be described as in Fig. 8. Let us consider a pair of electrons of the Fermi liquid moving in the $(0,\pi)$ direction, i. e. they are in $\Gamma M$ direction of the Fermi surface of Bi2212 [21], with momentum -k and +k as shown in Fig. 8.

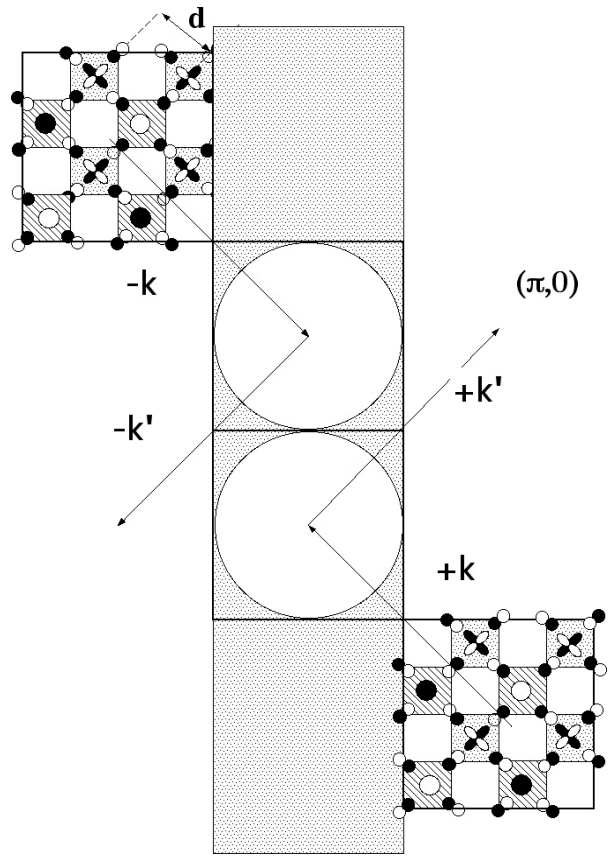

**Fig. 8.** Pictorial view of the pairing mechanism involving pairing of electrons in the Fermi surface moving in the $(0\pi)$ direction mediated by a pair of $3d_{3z^2-r^2}$ polarons. The polarons have condensed at $T^*>T_c$ into a Wigner one-dimensional charge density wave, forming stripes indicated by the shaded region. The symmetry of the wave function of the electrons moving in the $(0\pi)$ direction is shown [from ref. 23] in order to stress the point that the electrons moving in this direction have a mixed character of $b_1$ and $a_1$ local symmetry.

The wave function of the electron moving in the $(0,\pi)$ direction has not a pure $b_1$ character (at the Cu site $3d_{x^2-y^2}$) as for the electrons moving in the $(\pi,\pi)$ direction. The wave function of the electron moving in this direction have a mixed $b_1$ and $a_1$ character ($3d_{3z^2-r^2}$ and $3d_{x^2-y^2}$) as calculated by Mahan [23] and it is shown in Fig.

8. The localized polarons have a large effective mass resulting from the lattice distortion which forms a molecular orbital with a similar mixture of $b_1$ and $a_1$ character. The mobile electron on the large Fermi surface in the ΓM direction is coupled below $T_C$ with the localized polaron in the small Fermi surface centered at the M point in hole doped superconductors.

The pairing between electrons in the Fermi liquid is mediated by the virtual polaron pairs condensed in a Wigner paired CDW. Therefore they will act as negative Hubbard centers for the Fermi liquid. Because the polaron gas and the Fermi liquid are coupled below $T_C$ a more correct description of the pairing mechanism that we propose here can be given.

*The superconducting phase results from the competition at $T<T_C$ with a microscopic phase separation between a generalized Wigner polaronic CDW and a Fermi liquid. This phase separation takes place at $T<T^*$ because of an instability of the 2D-electron gas at densities $\rho_c/2 > \rho > \rho_c$ and $\rho_c > \rho > 2\rho_c$ close to the critical density for a generalized Wigner polaron crystal.*

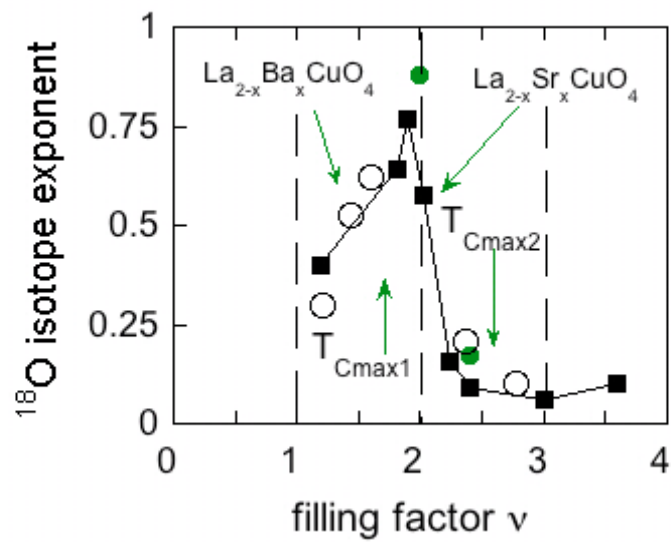

**Fig. 9.** The isotope exponent for LaBaCuO (open circles) and LaSrCuO (filled squares) systems from ref. [5]. In low doping regime, $1<\nu<2$, the two electrons of the pair remain in two different stripes (see Fig. 8) therefore the Coulomb repulsion is close to zero and the isotope effect is close to 0.5.

Now we want to show that this pairing model for the cuprate superconductors is supported by the experiments on the isotope effect. The isotope effect is described by the oxygen isotope exponent α measured by the shift in transition temperature where $T_c$= const $M^{-\alpha}$. The experimental investigations now converge toward the fact the isotope exponent in most of the families of high $T_c$ superconductors is in the range 0.5 -1 in low doping regime while it is very small in the high doping regime.

The common interpretation for this variation is that the isotope exponent decreases with increasing the critical temperature. The data however do not support this interpretation because it predicts a smooth variation of α with $T_c$ that is not observed. We have plotted in Fig. 9 the isotope exponent for LaBaCuO and LaSrCuO systems as function of the polaron pair filling factor. We show that the isotope exponent is 0.5 - 1 where $\delta_i < \delta_l$ and it is constant at about 0.1 where $\delta_i > \delta_l$. These data strongly support the pairing mechanism proposed here. In the low doping regime 1<ν<2 where $\delta_i < \delta_l$, where the superconducting quantum state with $T_{c1max}$ is realized for $<\delta_i>/<\delta_l>= 1$, the effective Coulomb repulsion μ* between the pairs is negligible. Therefore in spite of the short coherence length the isotope exponent is close to 0.5 like in the weak coupling regime for standard BCS superconductors where μ* is negligible because of the very large coherence length. In fact the Coulomb repulsion for the charges condensed into the Wigner polaronic CDW is compensated by the crystallization.

The mobile pairs of the Fermi liquid, in the pairing mechanism described in Fig. 8, are located in two different stripes and therefore the Coulomb repulsion is negligible. The sharp drop of the isotope exponent to about 0.1 takes place where $\delta_i > \delta_l$. In this regime 2<ν<4, where the superconducting quantum state with $T_{c2max}$ is realized for $<\delta_i>/<\delta_l>=2$, the density of the Fermi liquid becomes high and the probability for the mobile pairs of the Fermi liquid to be located in the same stripe becomes relevant. Now we want to show that the proposed pairing mechanism correctly predicts the Uemura plot [6].

A two dimensional (2D) Fermi liquid, of fixed density ρ, can be driven to the maximum $T_c$ by driving the electron gas close to the Wigner localization limit, where each electron is confined inside a sphere of diameter $d=2r_s a_B$. The Pippard coherence length of an eventual superconducting phase will be $\xi_0 =2r_s a_B$. For a 2D Fermi liquid, $k_F=R\ 2\pi\rho$, in this quantum state the critical temperature will depend only on ρ/m*

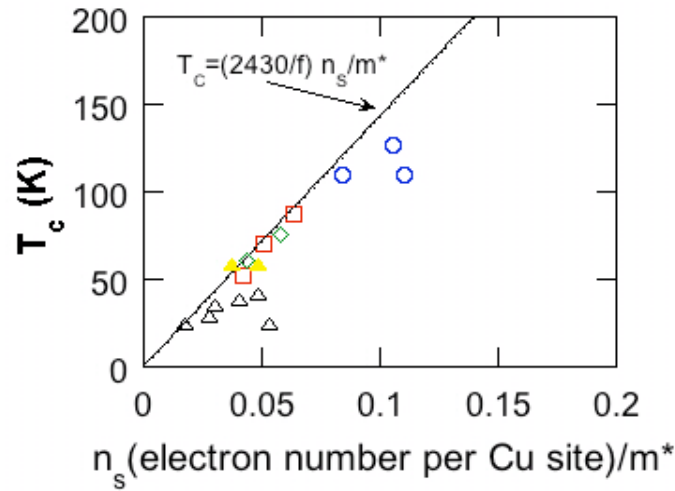

**Fig. 10.** The solid line is the calculated curve of $T_c$ versus $n_s/m^*$ for a condensate formed via the pairing mechanism described here $T_c$ (K) = 2430/f ($n_s/m^*$) by using f=1.7, valid for Bi2212. The theoretical prediction is compared with the data of $T_c$ versus $n_s/m^*$ as measured by muon spin relaxation experiments by Uemura et al. (empty simbols) and Keller et al. (filled simbols) for different materials (triangles La124; diamonds Y123 (ortho II phase); squares Bi2212; circles Bi2223).

$$K_B T_c = \frac{\hbar^2}{2m_0} \frac{R^2 e^\gamma}{\pi} \frac{\rho}{f m^*}$$

where f is a phenomenological measure of the deviation from the weak coupling limit 2 $\Delta_0/ K_B T_c$= 3.52 f, $\gamma$ is the Euler's constant 0.577 and $R^2 e^\gamma /\pi$=0.8. If $\rho$ is measured in Å$^{-2}$, $T_c(K) = 35440.8 \frac{\rho}{f m^*}$ In our model of high $T_c$ superconductivity *only the portion of the Fermi liquid coupled with the paired CDW contributes to the condensate. Therefore $T_c$ follows the formula given above where we have to insert as condensate density* $\rho_s= 2n_s/a^2$, where $n_s$ is the number of electrons per Cu site. We obtain an expression for the cuprates that can be compared with the experiments:

$$T_c(K) = 2430 \frac{n_s}{f m^*}$$

Where m* is the effective mass for the Fermi liquid. In order to calculate the predicted $T_c$ for a particular family we have to measure the ratio $2\Delta_0/K_B T_c$ to get f. In the Bi2212 family $2\Delta_0/K_B T_c$ has been found to be around 6 giving f=1.7. By using this empirical factor we can quantitatively predict the so called "Uemura plot" of $T_c$ versus $n_s/m^*$ as measured by muon-spin relaxation rate [6]. The predicted relation will be different for different families because it depends on the fenomenological parameter f, which measures the deviation from the weak coupling limit and which can also vary with doping. The comparison between the calculated curve and the data for Bi2212 in Fig. 10 is very good. The deviation from the calculated curve at high doping and for different families can be assigned to the variation of f.

In conclusion we have shown that the measure of the polaron size sheds a new ligth on many physical properties of cuprate superconductors, and indicates a possible solution for the microscopic pairing mechanism in high $T_c$ superconductors.